# One-Dimensional Numerical Study on Ignition of the Helium Envelope in Dynamical Accretion during the Double-Degenerate Merger

Kazuya Iwata[1,2,3], Keiichi Maeda[3]

## Abstract

In order for a double-detonation model to be viable for normal type Ia supernovae, the adverse impact of helium-burning ash on early-time observables has to be avoided, which requires that the helium envelope mass should be at most 0.02 M⊙. Most of the previous studies introduced detonation by artificial hot spots, and therefore the robustness of the spontaneous helium detonation remains uncertain. In the present work, we conduct a self-consistent hydrodynamic study on the spontaneous ignition of the helium envelope in the context of the double-degenerate channel, by applying an idealized one-dimensional model and a simplified 7 isotope reaction network. We explore a wide range of the progenitor conditions, and demonstrate that the chance of direct initiation of detonation is limited. Especially, the spontaneous detonation requires the primary envelope mass of ≳ 0.03 M⊙. Ignition as deflagration is instead far more likely, which is feasible for the lower envelope mass down to ~ 0.01 M⊙, which might lead to subsequent detonation once the deflagration to detonation transition (DDT) would be realized. High-resolution multi-dimensional simulations are required to further investigate the DDT possibility, as well as accurately derive the threshold between the spontaneous detonation and deflagration ignition regimes. Another interesting finding is the effect of the composition; while mixing with the core material enhances detonation as previously suggested, it rather narrows the chance for deflagration due to the slower rate of $^{12}C(\alpha,\gamma)^{16}O$ reaction at the lower temperature ~$10^8$K, with the caveat that we presently neglect the proton-catalyzed reaction sequence of $^{12}C(p,\gamma)^{13}O(\alpha,p)^{16}O$.

*Keywords:* Type Ia supernovae (1728), Hydrodynamical simulations (767)

---

[1] Corresponding author iwata.kazuya.6f@kyoto-u.ac.jp

[2] Department of Mechanical Engineering and Science, Kyoto University, Kyoto daigaku-Katsura, Nishikyo-ku, Kyoto 615-8540, Japan

[3] Department of Astronomy, Kyoto University, Kitashirakawa-Oiwake-cho, Sakyo-ku, Kyoto 606-8502, Japan

# 1. Introduction

Type Ia supernovae (SNe Ia) had long been believed to originate from uniform systems, as inferred by their one-parameter nature of observational properties, e.g., a relation between the peak luminosity and the evolution timescale (i.e., the so-called Phillips relationship which forms a basis for their use as cosmological standard candles) (Phillips 1993). However, despite the long-term debate, a conclusion has not yet been reached on the progenitor systems (Branch et al. 1995). Indeed, an increasing sample has led to discoveries of various subclasses and outliers with much more diverse natures (e.g., Taubenberger 2017) than previously known (Filippenko et al. 1992a, Filippenko et al 1992b). With the huge diversities seen in different SN Ia subclasses, the possibility that SNe Ia are indeed a mixture of several populations, associated with different progenitor systems and explosion mechanisms, has been seriously considered (Maeda & Terada 2016).

One of the main open problems regarding the nature of SNe Ia is this; what are the primary progenitor system and the explosion mechanism of normal SNe Ia? A white dwarf (WD) near the Chandrasekhar mass (Whelan & Iben 1973, Nomoto 1982a) has been insensibly studied as a promising progenitor. However, the single-degenerate (SD) binary system (Whelan & Iben 1973, Nomoto 1992), i.e., an accreting WD from a non-degenerate companion star, suffers from a lack of direct observational evidence of the non-degenerate companion star for normal SNe Ia (see, e.g., Maeda & Terada 2016, for a review). Another popular scenario, double-degenerate (DD) binary system (a merger of binary WDs; Iben & Tutukov 1984, Webbink 1984), on the other hand has a theoretical problem; it likely results in the formation of an ONeMg WD before reaching the Chandrasekhar mass, which does not explode as an SN Ia (Saio & Nomoto 1985, Schwab 2021).

As an alternative scenario, the double-detonation mechanism on a sub-Chandrasekhar-mass WD, which could be realized both in the SD and DD systems, has attracted attention from the community in the last decade. In this scenario, detonation initiated in the helium envelope of the accreting WD subsequently induces secondary detonation, either at the center of its carbon/oxygen core (Nomoto 1982b, Livne 1990), in the core but well away from the center (Fink et al. 2010, Boos et al. 2021 Fig. 3), or at the interface between the core and the envelope (Livne & Glasner 1990, Gronow et al. 2021). There is (at least) a peculiar subclass of SNe Ia that has been suggested to be triggered by the double-detonation mechanism (Jiang et al. 2017, Kupfer et al. 2021, De et al 2019). Also, one of the hyper-velocity WDs recently discovered could be a smoking gun of the double-detonation channel in the DD merger (Shen et al. 2018).

A key issue in the double-detonation scenario is the following; it has to avoid the adverse impact of helium-burning ash on early-time observables to be a feasible model for normal SNe Ia. Previous studies on synthetic spectra and light curves (Shen et al. 2021, Woosley and Kasen 2011, Maeda et al. 2018) demonstrated that envelope mass of $> 0.02$ $M_\odot$ should leave a trace of the helium-burning ash in observables, which disagrees with the observations of normal SNe Ia. Hence, the robustness of the double-detonation mechanism with a little amount of the helium envelope ($< 0.02$ $M_\odot$) has to be explored.

Recent several multidimensional hydrodynamic simulations (Boos et al. 2021, Townsley et al. 2019a, Fink et al. 2010) succeeded in exploding a whole WD by the double-detonation mechanism with the primary helium envelope of ≲ 0.02 M☉. However, this is not a direct proof for the detonation mechanism; detonation in the envelope in most of these studies was ignited with the use of artificial hot spots. One of a few studies conducting self-consistent simulations of spontaneous ignition (Guillochon et al. 2011) found the formation of a hot spot induced by the Kelvin-Helmholtz instability between the accretion stream and the relatively massive accretion torus (≳ 0.05 M☉), which subsequently led to detonation. However, a strong resolution dependence was indicated in one of the successful cases, where the detonation originally found in the lower-resolution simulation turns out to fail in the higher-resolution simulation (with a resolution of 18 km). Pakmor et al. (2013) also demonstrated that spontaneous ignition of helium detonation was realized in a 0.01 M☉ envelope with a resolution of ~10km. Later, Pakmor et al. (2021) found helium detonation on the envelope of even ~$10^{-3}$-$10^{-2}$ M☉, using a cell size of ~15 km resolved in the hot spot, although their model resulted in the core detonation of the secondary rather than the primary, which may be applicable only for a faint and rare subclass of SNe Ia.

As summarized above, the results of the previous studies have not yet been converged on the robustness of spontaneous detonation. Furthermore, since all these studies are whole star simulations, the high computation cost limits the investigation; (1) the parameters surveyed to date are still limited, and (2) the applied resolutions, ~10km, may not be necessarily sufficient to capture the ignition process. Hence, the robustness and conditions for spontaneous ignition of helium detonation are still uncertain, which has motivated us to explore a wide range of the progenitor conditions favorable for the helium ignition; the primary mass, the envelope mass, its composition, and the accretion rate from the companion WD. In the present work, we apply a simplified/idealized numerical model in one dimension as the initial stage of the survey; this allows us to study a large parameter space, as well as to achieve a better resolution than the previous studies.

## 2. Numerical setup

Fig. 1 summarizes the numerical setup for our spherically symmetric one-dimensional simulation. A WD is composed of a carbon/oxygen core, and a helium-rich envelope surrounding the core. The center of the core is located at the left symmetric boundary ($r$=0). The hydrostatic profile of the WD is derived on the assumption of a constant temperature of $3\times10^7$K for the core and an adiabatic temperature profile for the envelope. The pressure balance at the interface between the core and the envelope is assured as a boundary condition in the construction of the initial profile. The mass of the core ($M_{core}$) is varied for a range of 0.8-1.1 M☉, and the core composition is fixed to be C:O =0.5:0.5 in the mass fractions.

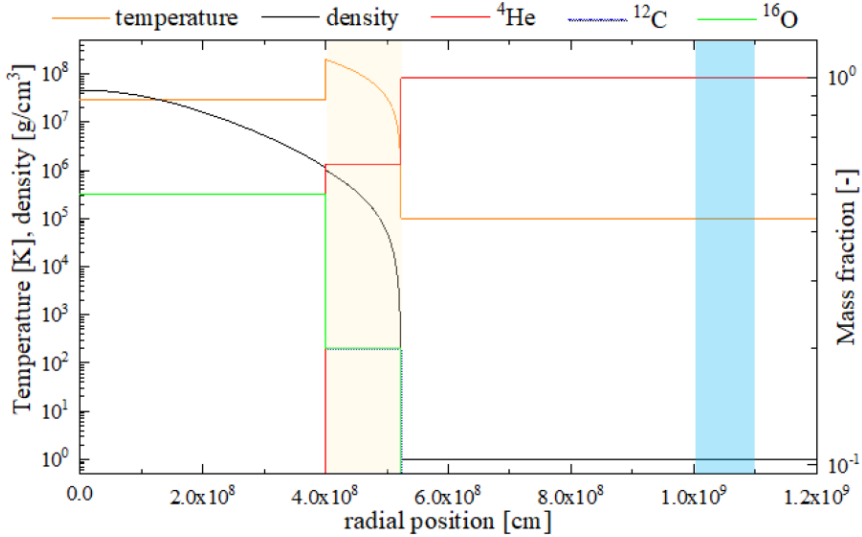

**Fig. 1** One-dimensional setup exemplified for the envelope with $T_{base}=2\times10^8$ K and the mixed-envelope composition (He:C:O=0.6:0.2:0.2 in the mass fractions). The envelope region is colored yellow and the region with the mass accretion source is colored blue.

The helium-rich envelope is varied in mass ($M_{env}$) between 0.01-0.05 M☉, and the envelope composition is set to be either a pure helium or a mixed composition with He:C:O=0.6:0.2:0.2 ('mixed envelope' hereafter). The mixed-envelope composition is chosen as guided by several previous studies (Shen & Moore 2014, Tanikawa et al. 2018), considering a mixing between the envelope and the core materials induced by the Kelvin-Helmholtz instability (Guillochon et al. 2011). We note that this mixing is suggested to greatly enhance detonation in the envelope (Shen & Moore 2014). The other parameter for the envelope is the base temperature, which is chosen to be either $T_{base} = 3\times10^7$ K (i.e., the cold envelope formed through the pre-merger evolution stage) or $T_{base} = 1-3\times10^8$ K (i.e., the hot envelope considering the accreted material in the merger process prior to the dynamical phase). A low-density, cold circumstellar material at 1g/cm³ and 1x10⁵K is placed around the WD; its properties are set so as not to affect the dynamical evolution of the accretion flow studied in the present work.

Pure-helium mass accretion from the companion star during the merger is represented as a mass source placed in a spherical shell between $r_1=1.0\times10^9$ cm and $r_2=1.1\times10^9$ cm. For simplicity, a mass source term at every cell (i.e., a shell with the inner and outer radii being $r_1$ and $r_2$) is given as $3a\dot{M}_{acc}/4\pi(r_2^3-r_1^3)$. The mass accretion rate $\dot{M}_{acc}$ is a parameter, which is set to be in the range of $2\times10^{-5} - 2\times10^{-3}$ M☉s⁻¹ (Dan et al. 2010). We introduce the geometrical factor, $a=10$, considering accretion in a disk-like form which is more likely to occur in the DD merger system (Tanikwa et al. 2015): the mass flux is increased from a spherically symmetric value by a factor of 10, reflecting the effective solid angle $2\pi h/R_{WD}$ covered by the accretion disk with the height of h ~ $10^8$ cm surrounding a typical WD radius $R_{WD}$ ~ $5\times10^8$cm. The initial velocity of the stream is set to be zero, and the temperature is set the same as the surrounding material (1x10$^5$ K). The total physical time for the simulation is ~ 45 s, which is sufficiently longer than the typical ignition timescale of ~10s (see, e.g., Pakmor et al. 2021).

Eulerian cells are distributed throughout the simulation domain. In our standard runs, the minimum cell size is 1 km to resolve the entire envelope. It is smaller than those applied in most of the relevant studies. This is only competed by the resolution of ~0.95-2.4 km applied by Glasner et al. (2018), who discussed spontaneous detonation for an accretion rate of ~10$^{-8}$ M⊙yr$^{-1}$ in the context of the conventional SD scenario. The cell size is smoothly expanded inward and outward, with the resolution downgraded to 10 km at the center of the WD and the region with the mass source term.

For the model set up as described above, Euler equations are solved with explicit time integration using the 3rd order TVD Runge Kutta scheme (Shu & Osher 1988). Advection term is discretized by the Ausm+up scheme which was originally developed as an all-speed flux scheme for subsonic to hypersonic velocity (Kitamura & Shima 2013), which has been applied to a number of aerospace hydrodynamic problems. Second-order accuracy in space is assured by applying a simple minmod-limiter for interpolation of the primary variables (Waerson and Deconinck 2007). Equation of state (EOS) by Timmes and Arnett (1999) for arbitrary degrees of degeneracy and relativity is used in a tabulated form. The same EOS is applied to provide the hydrostatic profile of the WD, hence no post-relaxation is required. Gravity on each cell is computed for the integrated mass from the center.

The alpha reaction network consisting of 7 isotopes (Timmes et al. 2000b) is applied, which treats the Intermediate Mass Elements (IME) and Iron Group Elements (IGE) as one unified element each, as represented by Si and Ni. This is simpler than the frequently used 13 isotope network, but should be enough for the present focus on the initial ignition stage of helium. The caveat is that two mutually-related effects are omitted in the present study; the existence of $^{14}$N, which could produce proton through the reaction sequence of $^{14}$N($\alpha,\gamma$)$^{18}$F($\alpha$,p)$^{21}$Ne, and the catalytic effect of proton through the reaction sequence of $^{12}$C(p,$\gamma$)$^{13}$N($\alpha$,p)$^{16}$O. The latter reaction has been reported to boost up the initiation of detonation in the envelope (Shen & Moore 2014). The contribution of $^{14}$N to produce proton in the lighter envelope mass < 0.1 M⊙ is a little uncertain since the lifetime of $^{18}$F is relatively longer until the moment of ignition even for the heavy envelope of 0.2M⊙ (Figure 8 in Shen & Bildsten 2009). Instead, preexisting proton with the mass fraction of 10$^{-4}$ would greatly affect the ignition behavior (Shen & Moore 2014), which we postpone for future work.

Actual temporal integration of the reaction network is implemented iteratively using the variable-order Bader-Deuflhard method (Bader and Deulfhard 1983). The operator-splitting scheme is applied to separately integrate the flow dynamics and the nuclear reactions.

## 3. Results and discussion

Successful ignition of the envelope is attained within the total duration of ~45s only for the base temperature of $T_{\text{base}}$=2x10$^8$ K, whereas no ignition is found to occur for $T_{\text{base}}$≤1x10$^8$ K (as is confirmed by our testing $T_{\text{base}}$=3x10$^7$K and 1x10$^8$ K). We find that prompt ignition occurs without the help of the accretion for $T_{\text{base}}$≥3x10$^8$

K. It is therefore indicated that the cold envelope (~$10^7$ K) likely obtained during the progenitor WD evolution stage is hard to ignite, while a favorable condition for helium ignition could be realized in the hot envelope formed during the merger where the virial temperature reaches ~$10^8$ K. Hence, in most of the parameter space we explore in the present work, the base temperature is fixed at $T_{base}=2\times10^8$ K. As for the accretion rate, we find that the accretion rate of $2\times10^{-5}$ M⊙s$^{-1}$ never achieves ignition in the simulations. It is thus indicated the accretion rate in the range of $10^{-4} - 10^{-3}$ M⊙s$^{-1}$ is needed for the dynamical ignition, and hence the two values are primarily explored for the accretion rate: $2\times10^{-4}$ and $2\times10^{-3}$ M⊙s$^{-1}$.

Each case is labeled like '10H5S'. '10' denotes a tenfold value of $M_{CO}$, 'H' means the pure helium envelope ('M' is used instead for the mixed envelope), '5' denotes a hundredfold value of $M_{env}$, and 'S' denotes the smaller accretion rate of $2\times10^{-4}$ M⊙s$^{-1}$ ('L' is used instead for the larger accretion rate of $2\times10^{-3}$ M⊙s$^{-1}$), respectively. Under this naming rule, case 10H5S is the case where $M_{CO}=1.0$M⊙ with the pure helium envelope of $M_{env}=0.05$M⊙, and with the accretion rate of $2\times10^{-4}$ M⊙s$^{-1}$. The details of the model setup for each case, along with the simulation results, are summarized in Table 1 and 2.

### 3.1 Ignition behaviors

Fig. 2 illustrates the temporal evolutions of the three regimes of ignition; detonation, shocked subsonic flame, and isobaric ignition. In all the cases, ignition is triggered locally at the base of the envelope (the left end of each panel), where the nuclear timescale is substantially shorter than the local dynamical timescale (for detonation) or they are comparable (for shocked subsonic flame). Detonation is developed rapidly, for which the flame front represented by a sudden temperature rise propagates at the velocity of ~$1\times10^9$ cm/s, being consistent with the Chapman-Jouguet velocity (Chapman 1899, Timmes & Niemeyer 2000c). In the regime of shocked subsonic flame, a weak shock precedes a sub-sonic burning front; the shock speed (~$2.4\times10^8$ cm s$^{-1}$) slightly exceeds the sound speed (~$2.1\times10^8$ cm s$^{-1}$), but the flame velocity is subsonic while being comparable to the sound speed. However, its propagation is different from that of deflagration, which is a subsonic flame driven by transport phenomena (Timmes 2000a), since the present simulation does not include heat conduction nor microscopic turbulence. It is rather driven by the shock-induced self-ignition as it is the case for detonation. From this perspective, the shocked subsonic flame in this study can be regarded as an intermediate ignition regime between detonation and deflagration. In terrestrial chemical systems, a similar intermediate flame regime is observed at the last stage of Deflagration-to-Detonation Transition (DDT), which is called 'fast flame' or 'high-speed deflagration' (Chue et al. 1993, Lee 2008). Nevertheless, propagation speed of fast flame is known to be around the sound speed of burned ash (~$7\times10^8$ cm s$^{-1}$ in this study), which is much higher than those of the shocked subsonic flames observed here ($6\times10^7$-$2\times10^8$ cm s$^{-1}$), partly due to a lack of the modeling of multidimensional convective and turbulent effects. Since these effects are also needed for the subsequent development of deflagration to detonation (Timmes & Niemeyer 2000c, Townsley et al. 2019), shocked subsonic flames in this study do not evolve into detonation in our simulations.

**Table 1** Summary of simulation results for the pure helium envelope

| Case | Variant | $M_{core}$ [$M_\odot$] | $M_{env}$ [$M_\odot$] | $X_{He}$ | $\dot{M}_{acc}$ [$10^{-4} M_\odot s^{-1}$] | $\rho_{b,ini}$ ($10^6$g/cm$^3$) | Limiter (dln$T_{max}$) | $\Delta r_{min}$ [km] | Burning Regime | $t_{ign}$ [s] |
|---|---|---|---|---|---|---|---|---|---|---|
| 10H5S | | 1.0 | 0.05 | 1.0 | 2 | 0.97 | off | 1 | B | 16 |
| 10H4S | | 1.0 | 0.04 | 1.0 | 2 | 0.76 | off | 1 | C | 18 |
| 10H3S | | 1.0 | 0.03 | 1.0 | 2 | 0.56 | off | 1 | C | 22 |
| 10H2S | | 1.0 | 0.02 | 1.0 | 2 | 0.37 | off | 1 | C | 28 |
| 10H1S | | 1.0 | 0.01 | 1.0 | 2 | 0.19 | off | 1 | C | 30 |
| 10H5L | | 1.0 | 0.05 | 1.0 | 20 | 0.97 | off | 1 | B | 8 |
| 10H4L | | 1.0 | 0.04 | 1.0 | 20 | 0.76 | off | 1 | B | 8 |
| 10H3L | | 1.0 | 0.03 | 1.0 | 20 | 0.56 | off | 1 | B | 9 |
| 10H2L | | 1.0 | 0.02 | 1.0 | 20 | 0.37 | off | 1 | C | 10 |
| 11H5S | | 1.1 | 0.05 | 1.0 | 2 | 1.71 | off | 1 | A | 6 |
| | 11H5Sl020 | 1.1 | 0.05 | 1.0 | 2 | | 0.20 | 1 | B | 6 |
| | 11H5Sc025 | 1.1 | 0.05 | 1.0 | 2 | | off | 0.25 | B | 6 |
| 11H4S | | 1.1 | 0.04 | 1.0 | 2 | 1.32 | off | 1 | C | 11 |
| | 11H4Sl020 | 1.1 | 0.04 | 1.0 | 2 | | 0.20 | 1 | C | 11 |
| 11H3S | | 1.1 | 0.03 | 1.0 | 2 | 0.97 | off | 1 | C | 16 |
| 11H2S | | 1.1 | 0.02 | 1.0 | 2 | 0.65 | off | 1 | C | 20 |
| 11H1S | | 1.1 | 0.01 | 1.0 | 2 | 0.34 | off | 1 | C | 23 |
| 11H5L | | 1.1 | 0.05 | 1.0 | 20 | 1.71 | off | 1 | A | 6 |
| | 11H5Ll020 | 1.1 | 0.05 | 1.0 | 20 | | 0.20 | 1 | B | 6 |
| | 11H5Ll025 | 1.1 | 0.05 | 1.0 | 20 | | 0.25 | 1 | A | 6 |
| | 11H5Lc025 | 1.1 | 0.05 | 1.0 | 20 | | off | 0.25 | B | 6 |
| 11H4L | | 1.1 | 0.04 | 1.0 | 20 | 1.32 | off | 1 | A | 7 |
| | 11H4Ll020 | 1.1 | 0.04 | 1.0 | 20 | | 0.20 | 1 | B | 7 |
| | 11H4Lc025 | 1.1 | 0.04 | 1.0 | 20 | | off | 0.25 | B | 7 |
| 11H3L | | 1.1 | 0.03 | 1.0 | 20 | 0.97 | off | 1 | B | 8 |
| 11H2L | | 1.1 | 0.02 | 1.0 | 20 | 0.65 | off | 1 | C | 8 |
| 09H5S | | 0.9 | 0.05 | 1.0 | 2 | 0.60 | off | 1 | C | 25 |
| 09H5L | | 0.9 | 0.05 | 1.0 | 20 | 0.60 | off | 1 | C | 10 |
| 08H5S | | 0.8 | 0.05 | 1.0 | 2 | 0.38 | off | 1 | C | 32 |
| 08H3S | | 0.8 | 0.03 | 1.0 | 2 | 0.21 | off | 1 | C | 38 |
| 08H1S | | 0.8 | 0.01 | 1.0 | 2 | 0.06 | off | 1 | C | 40 |

Notes: The character for the "Burning regime" denotes the following; (A) detonation; (B) shocked subsonic flame; (C) isobaric ignition; (D) no ignition.

Table 2 Summary of simulation results for the mixed envelope

| Case | Variant | $M_{core}$ [$M_\odot$] | $M_{env}$ [$M_\odot$] | $X_{He}$ | $\dot{M}_{acc}$ [$10^{-4} M_\odot s^{-1}$] | $\rho_{b6,ini}$ ($10^6 g/cm^3$) | Limiter ($dlnT_{max}$) | $\Delta r_{min}$ [km] | Burning Regime | $t_{ign}$ [s] |
|---|---|---|---|---|---|---|---|---|---|---|
| 10M5S | | 1.0 | 0.05 | 0.6 | 2 | 1.01 | off | 1 | B | 30 |
| 10M4S | | 1.0 | 0.04 | 0.6 | 2 | 0.79 | off | 1 | D | - |
| 10M5L | | 1.0 | 0.05 | 0.6 | 20 | 1.01 | off | 1 | A | 11 |
| | 10M5Ll020 | 1.0 | 0.05 | 0.6 | 20 | | 0.20 | 1 | B | 11 |
| | 10M5Lc050 | 1.0 | 0.05 | 0.6 | 20 | | off | 0.50 | B | 10 |
| | 10M5Lc025 | 1.0 | 0.05 | 0.6 | 20 | | off | 0.25 | B | 10 |
| 10M4L | | 1.0 | 0.04 | 0.6 | 20 | 0.79 | off | 1 | B | 15 |
| 10M3L | | 1.0 | 0.03 | 0.6 | 20 | 0.59 | off | 1 | C | 16 |
| | 10M3Lc050 | 1.0 | 0.03 | 0.6 | 20 | | off | 0.50 | C | 16 |
| 11M5S | | 1.1 | 0.05 | 0.6 | 2 | 1.76 | off | 1 | A | 15 |
| | 11M5Sl020 | 1.1 | 0.05 | 0.6 | 2 | | 0.20 | 1 | B | 15 |
| | 11M5Sc025 | 1.1 | 0.05 | 0.6 | 2 | | off | 0.25 | B | 15 |
| 11M4S | | 1.1 | 0.04 | 0.6 | 2 | 1.36 | off | 1 | A | 20 |
| | 11M4Sl020 | 1.1 | 0.04 | 0.6 | 2 | | 0.20 | 1 | B | 20 |
| | 11M4Sc025 | 1.1 | 0.04 | 0.6 | 2 | | off | 0.25 | B | 20 |
| 11M3S | | 1.1 | 0.03 | 0.6 | 2 | 1.01 | off | 1 | B | 26 |
| | 11M3Sc200 | 1.1 | 0.03 | 0.6 | 2 | | off | 2 | A | 28 |
| 11M2S | | 1.1 | 0.02 | 0.6 | 2 | 0.67 | off | 1 | D | - |
| 11M5L | | 1.1 | 0.05 | 0.6 | 20 | 1.76 | off | 1 | A | 8 |
| | 11M5Ll010 | 1.1 | 0.05 | 0.6 | 20 | | 0.10 | 1 | B | 8 |
| | 11M5Ll020 | 1.1 | 0.05 | 0.6 | 20 | | 0.20 | 1 | A | 8 |
| | 11M5Ll025 | 1.1 | 0.05 | 0.6 | 20 | | 0.20 | 1 | A | 8 |
| | 11M5Lc050 | 1.1 | 0.05 | 0.6 | 20 | | off | 0.50 | A | 8 |
| | 11M5Lc025 | 1.1 | 0.05 | 0.6 | 20 | | off | 0.25 | A | 8 |
| 11M4L | | 1.1 | 0.04 | 0.6 | 20 | 1.36 | off | 1 | A | 9 |
| | 11M4Ll020 | 1.1 | 0.04 | 0.6 | 20 | | 0.20 | 1 | B | 9 |
| | 11M4Ll025 | 1.1 | 0.04 | 0.6 | 20 | | 0.25 | 1 | A | 9 |
| | 11M4Lc025 | 1.1 | 0.04 | 0.6 | 20 | | off | 0.25 | B | 9 |
| 11M3L | | 1.1 | 0.03 | 0.6 | 20 | 1.01 | off | 1 | A | 15 |
| | 11M3Ll020 | 1.1 | 0.03 | 0.6 | 20 | | 0.20 | 1 | B | 15 |
| | 11M3Lc025 | 1.1 | 0.03 | 0.6 | 20 | | off | 0.25 | B | 15 |
| 11M2L | | 1.1 | 0.03 | 0.6 | 20 | 0.67 | off | 1 | D | - |
| 09M5S | | 0.9 | 0.05 | 0.6 | 2 | 0.63 | off | 1 | C | 45 |
| 09M4S | | 0.9 | 0.04 | 0.6 | 2 | 0.49 | off | 1 | D | - |
| 09M5L | | 0.9 | 0.05 | 0.6 | 20 | 0.63 | off | 1 | C | 13 |
| 09M4L | | 0.9 | 0.04 | 0.6 | 20 | 0.49 | off | 1 | C | 14 |
| 09M3L | | 0.9 | 0.03 | 0.6 | 20 | 0.37 | off | 1 | C | 16 |
| 08M5S | | 0.8 | 0.05 | 0.6 | 2 | 0.41 | off | 1 | D | - |
| 08M5L | | 0.8 | 0.05 | 0.6 | 20 | 0.41 | off | 1 | C | 13 |
| 08M4L | | 0.8 | 0.04 | 0.6 | 20 | 0.32 | off | 1 | C | 14 |
| 08M3L | | 0.8 | 0.04 | 0.6 | 20 | 0.23 | off | 1 | D | - |

Both detonation and shocked subsonic flame observed in the simulations incinerate helium in the envelope and propagate outward, whereas isobaric ignition is confined locally at the base. The pressure profile is almost unchanged through the ignition. Velocity change is also minor compared to that in the shocked subsonic flame regime. A sudden decrease in density occurs in accordance with the increase in temperature to balance the surrounding pressure. Delayed ignition around the ignition spot (partly due to numerical diffusion) extends the flame region as the time proceeds, but it affects the surrounding materials very little since neither heat conduction nor convective/turbulent effects are included. This is why most of the mass in the envelope remains unburned till the end of the simulations. This ignition regime is expected to subsequently induce the propagation of deflagration, when heat conduction is included and the resolution is increased by several orders; helium deflagration in the envelope base is $< 10^3$ cm/s in the burning velocity and $< 10^5$ cm in the flame width (Timmes & Niemeyer 2000).

Fig. 3 shows the evolution paths of the envelope base through the accretion and ignition. The case 11H4S and 11H4L starts from the same initial condition with different accretion rates. The case 11H4S results in isobaric ignition, whereas the case 11H4L results in detonation. As shown here, the base density in both cases suddenly drops just after the ignition due to the expansion of the hot ash. For 11H4S, the base condition is eventually stabilized as the pressure balance is reached. For 11H4L, detonation is subsequently ignited, and the inward shock once compresses the base, which is then followed by the expansion as the outer detonation propagates further outward. This difference in the ignition regime due to the accretion rate can be attributed to the difference in their evolution paths prior to ignition, which is shown in the left panel of Fig. 3. Differently from the SD channel where the accretion rate is the order of $10^{-8}$ $M_\odot yr^{-1}$, hydrostatic assumption is not satisfied and the dynamical accretion compresses the envelope differently depending on the accretion rate, sometimes discontinuously via shock waves. As a result, while 11H4S experiences only a mild compression before ignition, the envelope base in 11H4L reaches much higher density.

As described above, in the context of a one-dimensional problem, isobaric ignition cannot trigger the double detonation, leaving most of helium unburned. Also, shocked subsonic flame will be too slow to generate a strong shock to detonate the WD core, since it will take too long (~10s) to wrap around the WD if ignition occurs on a point.

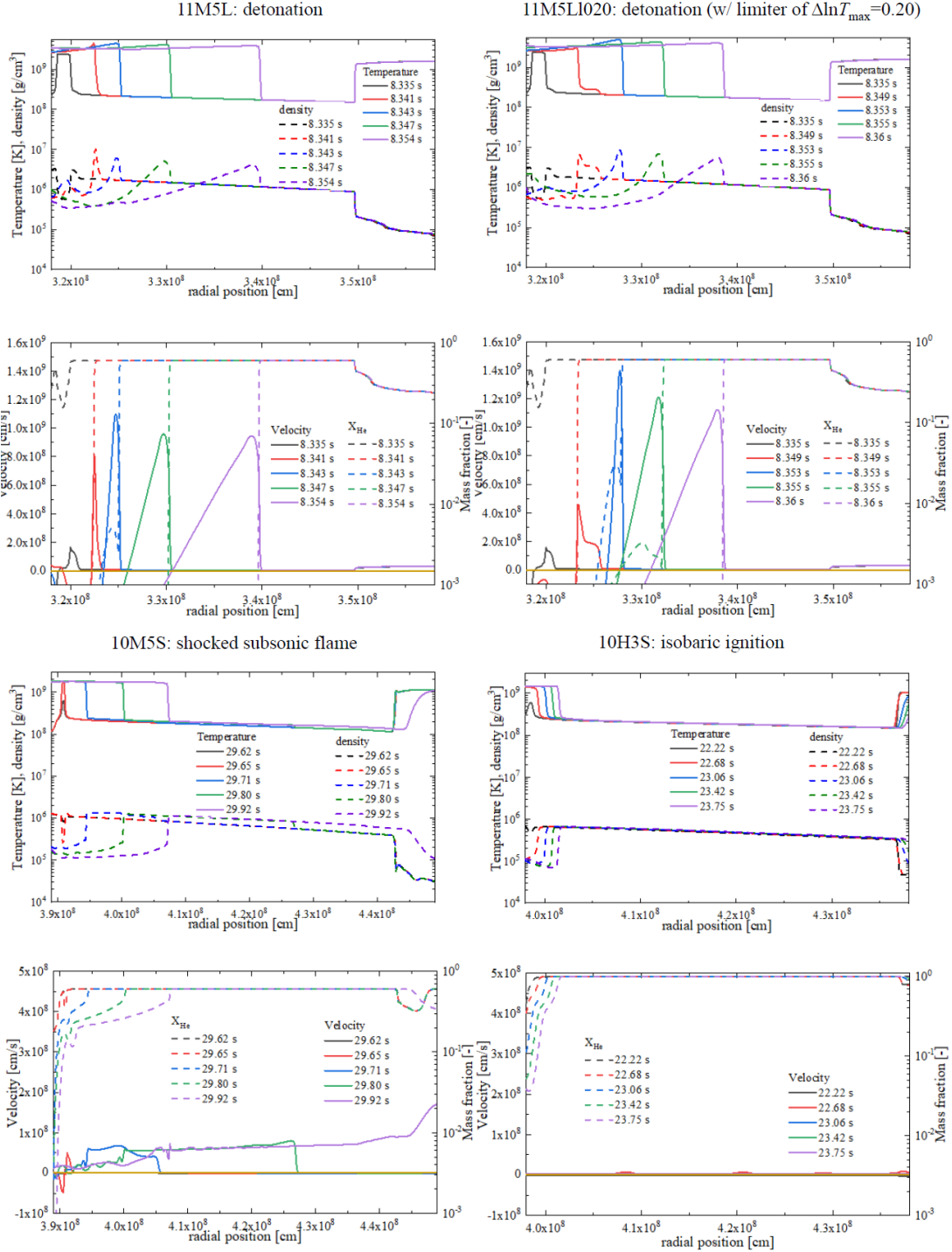

**Fig. 2** Temporal evolutions of the systems in the three burning regimes. Two separate figures are provided for each regime to show the profiles of temperature, density, velocity and helium mass fraction for five different instants, which are distinguished by the colors of the lines. For the detonation regime, two examples are shown with different treatments of the nuclear reactions (i.e., the burning limiter as explained in Section 3.4; upper two panels). The envelope base is located at the left end, and the accretion stream is partly shown close to the right end.

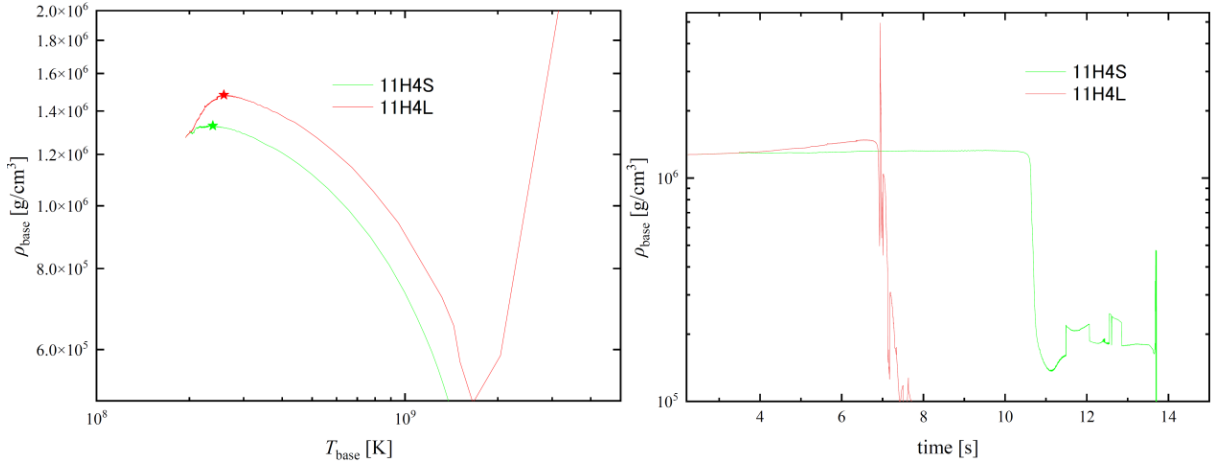

**Fig. 3** The left panel shows the evolution paths of the temperature/density of the envelope base for the case 11H4S (green) and 11H4L (red). Ignition condition is denoted by a star symbol. The right panel shows the temporal profiles of the base density for the two cases.

We however note that the fate of such a system, failing to ignite the spontaneous detonation, will require further investigation on how deflagration is subsequently developed and if DDT is achievable following propagation of the deflagration within the envelope. This DDT process is a requisite in delayed-detonation model for near-Mch scenario(Khokhlov 1991, Arnett and Livne 1994), but has not been considered in the context of the double-detonation scenario previously.

### 3.2 Progenitor conditions for ignition

Our simulation results with $T_{base}=2.0 \times 10^8$ K are summarized in Tables 1 and 2, for the pure helium envelope and the mixed envelope, respectively. Variants of the models in terms of numerical treatments are indicated by additional suffixes in the model names, like 'c025', which is for the case where the different minimum cell size of $\Delta r_{min}=0.25$ km is applied (a character "l" instead of "c" is also used for the cases using a burning limiter; see Section 3.4 for further details). Initial density at the base of the envelope $\rho_{b,ini}$ for each case is also shown in each table.

Burning regimes acquired for the range of the parameters are graphically illustrated in Fig. 4. Each panel corresponds to different combinations of the envelope composition and the accretion rate. We find that in the range of the progenitor/accreting conditions explored in the present work, the chance for direct ignition of detonation is limited to the envelope mass of $M_{env} \geq 0.03$ M⊙. Subsonic ignition is far more prevalent throughout the range of our survey, and it is found that the lighter WDs with $M_{CO}=0.8, 0.9$ M⊙ always result in isobaric ignition. Two distinct regimes of subsonic ignition are observed as stated earlier; shocked subsonic flame and isobaric ignition. The successful detonation is mainly limited to the heaviest core mass of $M_{CO}=1.1$ M⊙. In the context of a one-dimensional problem, it is thus indicated that direct ignition of detonation on a helium envelope with ≲ 0.02 M☉ (to avoid conflicts to observational properties; Shen et al. 2021, Woosley and Kasen 2011) is severely difficult for the progenitors of $M_{CO}$=0.9-1.0 M☉, which are favorable for normal type SNe Ia (Shigeyama et al. 1992, Sim et al. 2010, Woosley & Kasen 2011, Leung et al. 2020).

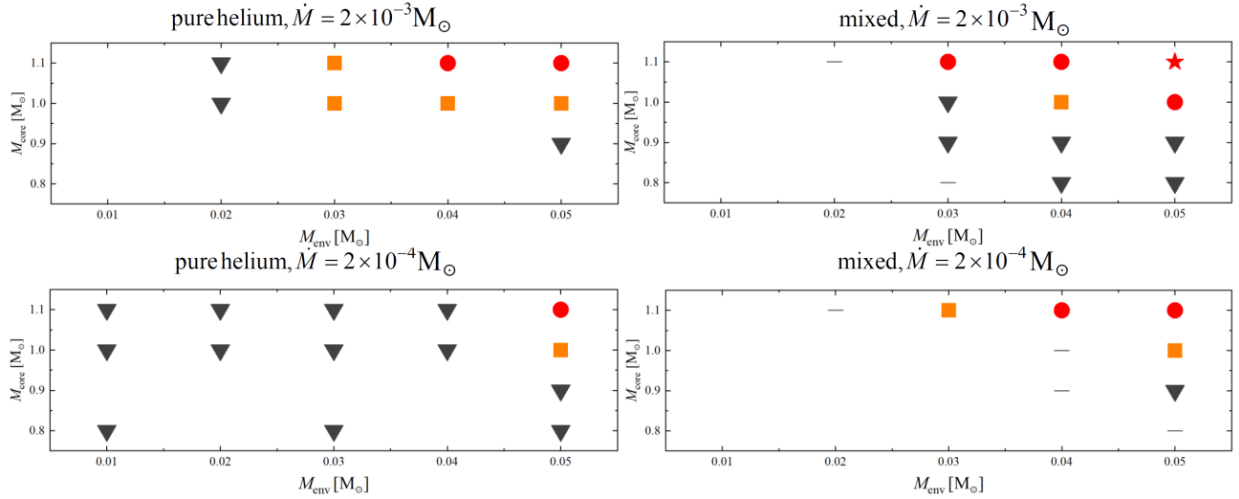

**Fig. 4** A graphical summary of the ignition regimes illustrated in the $M_{core}$-$M_{env}$ plane. It is shown separately for four different combinations of the composition of the primary envelope and the accretion rate. There are five ignition regimes observed; detonation (a red star), shocked subsonic flame (orange squares), isobaric ignition (black triangles), no ignition (gray bars), and the cases in which detonation occurs without the limiter but shocked subsonic flame is obtained instead with the limiter of $\Delta \ln T_{max}=0.20$ (red circles).

### 3.3 Effects of core-material mixing

We also find that the envelope composition plays an interesting role in the ignition process. The case 10M5L with the mixed envelope is an exception in that the lighter core of $M_{CO}$=1.0 M☉ succeeds in igniting detonation. The detonation is thus enhanced in the mixed envelope, as proposed by Shen & Moore (2014), who attributed this difference to the dominance of the reaction $^{12}C(\alpha,\gamma)^{16}O$ for higher temperatures of ≳ $10^9$ K.

However, in contrast to this, there is a far broader chance for subsonic ignition in the pure helium envelope than in the mixed envelope; 0.01 M☉ pure helium envelope on the lightest core WD succeeds in isobaric ignition (08H1S), while the mixed envelope achieves no ignition for $M_{env}$≲0.02 M☉. Carbon is not consumed in the `no-ignition' cases during the simulation time. Ignition time $t_{ign}$ is also longer for the mixed envelope. This is caused by a relatively lower temperature at the moment of ignition (~$10^8$ K), for which the rate of $^{12}C(\alpha,\gamma)^{16}O$ is much slower than the triple-alpha reaction (Shen & Moore 2014). Hence, subsonic flame/deflagration favors a larger content of helium. Recent studies on the double-detonation scenario have not paid attention to this point, since most of them introduce (by hands) hot spots of an already high temperature of ~$10^9$K. However, this aspect should be considered deliberately when realistic multi-dimensional simulations of the double-detonation model are implemented, since a subsonic ignition may lead to DDT, which could increase the chance for the double-detonation to occur. As shown here, the envelope composition is an important factor involved in such investigation. However, the caveat is that, as already stated (Section 2), the simplicity of the nuclear reaction network applied in the present work should be reassessed in the future by including proton-catalyzed α capture reactions and considering preexisting $^{14}$N/proton in the envelope, which may change the dependence on the envelope composition.

### 3.4 Resolution dependence and burning limiter

In the discussion on the spontaneous detonation, resolution dependence should be carefully assessed, since a spurious detonation can take place when the spatial resolution is not sufficient. The present study adopts $\Delta r_{min}$= 1km in the standard runs, which is among the highest resolutions applied in the studies on the spontaneous detonation. Note that we find that the conditions for the spontaneous detonation are very limited, already with this resolution. However, an even better resolution could differ in the prediction, especially on the threshold of detonation which may be numerically induced when the nuclear reaction timescale is not sufficiently resolved in the simulations (i.e., when $\tau_{nuc} < \Delta r/c_s$).

To avoid this problem, a burning limiter was proposed by Kushnir and Katz (2013), in which the nuclear reaction rate is suppressed so that the condition of $\tau_{nuc} > \Delta r/c_s$ is always satisfied. The prescription has been applied for resolving the detonation structure on a coarse cell (Kushnir and Katz 2019). This burning limiter has also been used in several studies on the spontaneous detonation in order to avoid a spurious detonation (Glasner et al. 2018, Pakmor et al. 2021). However, we have to be careful to configure the limiter and interpret its outcomes, since it tends to predict a lower propagation speed of the detonation front and may also adversely affect the igniting behavior including the time for ignition, although the latter problem was confirmed to be almost negligible by Pakmor et al. (2021) for a specific case.

To address the issues of the burning limiter and resolution dependence, we perform the simulations adopting the burning limiter with different configurations. We also test different resolutions, with the minimum cell sizes of 2.0, 0.5, and 0.25 km in addition to 1.0 km adopted in our standard runs. In the implementation of the burning limiter, the nuclear energy release rate in one time step is truncated so that the change of temperature, $\Delta \ln T$, does not exceed a prescribed upper limit $\Delta \ln T_{max}$ (Boos et al. 2021). In the present study, we investigate the cases with $\Delta \ln T_{max}$ =0.10, 0.20, and 0.25.

The simulations with different cell sizes and those with the burning limiter are denoted with a suffixe 'c' and 'l' in the model names, respectively, in Tables 1 and 2. The results with $\Delta r_{min}$= 0.25, 0.50 km are not different in the cases where shocked subsonic flame is observed with $\Delta r_{min}$= 1km; namely, shocked subsonic flame is observed irrespective of the spatial resolution. For the cases where detonation is observed with $\Delta r_{min}$= 1km, we find that the detonation is generally failed and shocked subsonic flame is instead observed for the higher-resolution runs; the only exceptions are the cases 11M5Lc025 and 11M5Lc050.

These results on the resolution dependence are also reproduced by the burning limiter of $\Delta \ln T_{max} =0.20$ with $\Delta r_{min}= 1$km, in a way that the burning limiter kills detonation and results in shocked subsonic flame, again except for the model sequence 11M5L. This is illustrated in Fig. 2; detonation is observed also in 11M5Ll020 (which is shown in Fig. 2 aligned with Model 11M5L), but it exhibits a delayed propagation at the initial stage of the ignition (t<8.35 s) at a smaller velocity ($\sim 4 \times 10^8$ cm/s). A similar behavior (i.e., a lower velocity with the limiter) is also seen in the previous studies (Glasner et al. 2018, Pakmor et al. 2021). For this specific model, however, in the later stage of the evolution it is seen that the detonation is accelerated to $\sim 1 \times 10^9$ cm/s, and the shock and the flame are strongly coupled, which looks not so different from the detonation observed without the limiter.

The choice of $\Delta \ln T_{max} =0.10$ results in a more pessimistic result in which detonation is never achieved, while for $\Delta \ln T_{max} =0.25$ the cases 11M4Ll025 and 11H5Ll025 also succeed in igniting detonation. This suggests that the detail of the implementation of the burning limiter can affect the outcome, and therefore ultimately it must be carefully tested with numerical resolution study as performed in the present work. To reproduce the outcomes of the most resolved simulations with $\Delta r_{min}= 0.25$km adopted in the present work, we infer that $\Delta \ln T_{max} = 0.20$ is the most appropriate choice that provides the resolution-convergent outcomes.

In any case, we conclude that the chance for the spontaneous detonation, which is severely limited already by our standard runs, becomes even lower when the resolution dependence and the burning limiter are considered. We find that the only condition in favor of the detonation based on the limiter with $\Delta \ln T_{max} =0.20$ is case 11M5L, which has the combination of the heaviest core and the heaviest envelope studied in the present work.

### 3.5 Comparison to the previous studies

How are our results justified against the previous studies which succeeded to initiate the helium detonation in the envelopes lighter than constrained in the present work? To address this issue, we first compare the ignition conditions found in the present study with those described by Woosley and Kasen (2011), which is based on spherical 1D numerical models. While Woosley and Kasen (2011) analyzed a different situation (the SD channel) from the present study, i.e., the thresholds for the occurrence of deflagration and detonation at the base of the pure helium envelope under a much lower accretion rate of $\sim 10^{-8}$ M$\odot$yr$^{-1}$ than explored in the present work, the comparison makes sense as long as we are interested in the local thermal conditions for the ignition of different burning regimes. Ignition temperature and density are defined as denoted by stars in Fig. 3, where the base density reaches the local maximum just before the sharp increase of temperature and the sharp decrease of density. Note that even in the detonation cases the base density initially drops at the moment of ignition due to the expansion of the hot ash, which is followed by the detonation trigger slightly away from the bottom of the envelope partly owing to numerical diffusion into the cold core material. In Fig. 5, the ignition conditions at the envelope base in the cases in our models (Tables 1 and 2) are represented by scattered plots, where different symbols are used for different burning regimes. Here, the filled and open symbols are for the pure helium envelope and the mixed one, respectively. Cases for shocked subsonic flames with/without the limiter of $\Delta \ln T_{max}=0.20$ are also shown by different symbols. The analytical thresholds for deflagration/detonation for the pure helium are denoted by the solid lines. The dashed line is the threshold for detonation arbitrarily configured by Woosley & Kasen (2011), for which they divided the threshold density for detonation by a factor of 4 to best reproduce the actual boundary found in their simulations between detonation and the other regimes. They attributed this discrepancy to several simplifications, e.g., the dimensionality of the problem (e.g., a point ignition instead of a spherical shell ignition) and the omission of gas dynamic effects.

Apparently, all the ignition conditions investigated in the present work are within the two analytical curves (for the pure helium composition), even including almost all the mixed envelope cases. It is seen that the shocked subsonic flame models are well below the detonation threshold and above the deflagration threshold, and the models resulting in isobaric ignition or no ignition are clustered around the deflagration threshold. Furthermore, Woosley & Kasen (2011) included $^{14}$N in the accreted material for the possible reaction sequence of $^{12}$C$(p,\gamma)^{13}$O$(\alpha,p)^{16}$O catalyzed by the proton produced through $^{14}$N$(\alpha,\gamma)^{18}$F$(\alpha,p)^{21}$Ne. Hence, this good agreement on ignition criteria reasonably supports the validity of our results, even though preexisting proton in the envelope may also enhance the ignition (Shen & Moore 2014).

It is also interesting that the dashed line delineates between isobaric ignition and the other regimes for the pure helium envelope reasonably well. For the mixed envelope, on the other hand, shocked subsonic flame is observed below this line, and the region for deflagration is narrower.

Woosley & Kasen (2011) obtained detonation over the dashed line, which is partly inconsistent with our results for the pure helium envelope. However, considering that shocked subsonic flame is a marginal regime close to detonation and that some shocked subsonic flame models denoted by circles turn into the detonation without the limiter, the agreement is not necessarily bad. We note that Woosley & Kasen (2011) did not use the limiter, hence the detailed treatment of the nuclear reactions (i.e., the limiter and the resolution) could alter their detonations into shocked subsonic flames as demonstrated in this study.

As for the lower limit for the envelope mass favorable for the detonation, Woosley & Kasen (2011) never succeeded in triggering the detonation for the envelope with the mass $< 0.02$ M$_\odot$: 0.0234 M$_\odot$ for 1.1 M$_\odot$ WD (model 11A), and 0.0315 M$_\odot$ for 1.0 M$_\odot$ WD (model 10A) are the lower limits for detonation in their models. The limits for the lighter WDs with $\leq 0.9$ M$_\odot$ are more severe in their simulations, which are around 0.05-0.06 M$_\odot$ in the envelope mass. If the burning limiter is introduced, the lower limit would be further increased. Therefore, our results in which the spontaneous detonation is failed for $M_{env} < 0.03$ M$_\odot$ (without the limiter) seem reasonable.

The next question is whether our results are consistent with the recent multidimensional studies on the double-detonation scenario in the DD mergers. Pakmor et al. (2013) and Guillochon et al. (2011) did not use the burning limiter. Guillochon et al. (2011) indicated the resolution dependence in which better resolutions failed in detonation. Strong resolution dependence is similarly confirmed in the present study, exemplified by case 11M3Sg200 which results in the spurious detonation with a coarser resolution of 2km than our standard run 11M3S (but we note that the resolution here is still better than those applied in the multi-dimensional models of Guillochon et al. 2011 and Pakmor et al. 2013).

A similar change would possibly happen to the detonation in the lighter envelope (0.01 M☉) observed by Pakmor et al. (2013) in their simulation, depending on the numerical methods chosen. Later, Pakmor et al. (2021) applied the burning limiter and still succeeded in the ignition of helium detonation. In their simulation, the envelope mass was $4\times10^{-3}$ M☉ prior to the dynamical accretion, and detonation occurred ~10 s later with the total accreted mass estimated to be ~0.01 M☉. The cells in the hotspot adopted in their simulation was ~15km. Their use of the burning limiter was configured so that the general outcome of the subsequent explosion of the secondary WD is the same as that of the simulation without the limiter. The Kelvin-Helmholtz instability disturbed and compressed the helium layer to create a hot spot with a temperature ~$10^9$ K and a density >$2\times10^5$ g/cm-3. This is far out of the range shown in Fig. 5, but a simple extrapolation of the dashed line to $10^9$K provides a density of $2.4\times10^5$ g/cm$^{-3}$ as the threshold for detonation. From one-dimensional perspective, hence, the occurrence of spontaneous detonation in their simulation is on a marginal border, and it can happen. This instability-induced compression cannot be reproduced by our one-dimensional simulation, and this implies multi-dimensional effects could enhance ignition. For further quantifying the thresholds for different burning regimes, we plan to extend our study to multi-dimensional configurations.

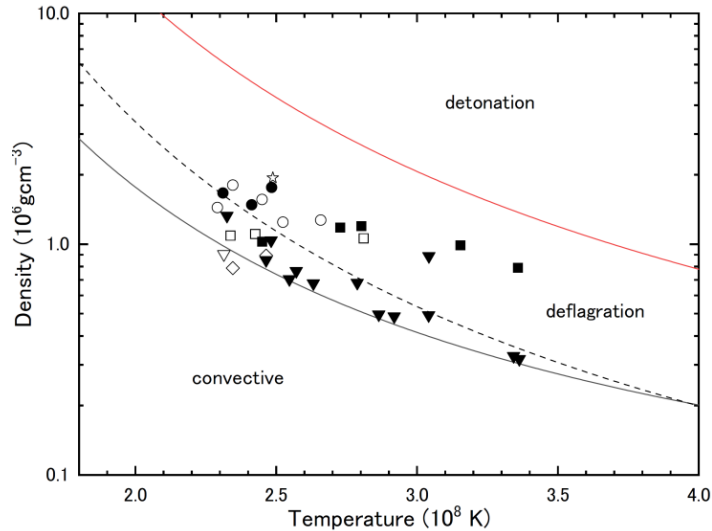

**Fig. 5** Ignition conditions as compared to the analytical thresholds for deflagration and detonation, for a pure helium envelope proposed by Woosley and Kasen (2011) (their Eqs. (8) and (7), respectively). The dashed line is the actual limit derived from their simulations between detonation and the other burning regimes. The symbols are for our models. Filled symbols denote the results for the pure helium envelope, and open ones are for the mixed envelope. Different symbols are used for the cases of detonation (star), shocked subsonic flame w/ limiter of $\Delta \ln T_{max}$=0.20 (circles), shocked subsonic flame w/o limiter (squares), isobaric ignition (triangles), and no ignition (diamonds).

## 4. Conclusions

To address an unresolved issue on prospects of a small amount of the WD envelope ($< 0.02$ M$_\odot$) to trigger the double detonation in the context of the DD merger, we have implemented a one-dimensional parametric study with a simple 7 isotope nuclear network to clarify the progenitor conditions favorable for the ignition of the envelope. Three different regimes of the thermonuclear ignition, i.e. detonation, shocked subsonic flame and isobaric ignition (which could lead to deflagration), are seen in the range of the parameters we explored. We find that the chance for the direct ignition of the detonation is very limited when the burning limiter and the resolution dependence are taken into account; the heaviest combination of 1.1 M$_\odot$ WD and 0.05 M$_\odot$ envelope with the composition He:C:O=0.6:0.2:0.2 could only lead to the detonation directly.

Enhancement of detonation induced by the contamination/mixing of the WD core material to the envelope (Shen & Moore 2014) is confirmed, but we have also found that the mixing rather decreases the chance for subsonic regimes of ignition (or thermonuclear nuclear ignition, in general); isobaric ignition is realized down to the envelope mass of 0.01 M$_\odot$ for the pure helium, but the envelope mass of $\gtrsim 0.03$ M$_\odot$ is required to ignite a flame for the mixed composition.

Therefore, in the context of one-dimensional problem, the spontaneous detonation within the envelope may require a possible but rather extreme combination of a massive WD, a massive envelope, and a high accretion rate. Especially problematic is the envelope mass range feasible for detonation $\gtrsim 0.03$ M$_\odot$ found in the present study, which does not satisfy the constraint from observational features of normal SNe Ia.

While our finding on the existence of different burning regimes depending on the progenitor conditions is qualitatively robust, our numerical study is still at an initial stage; the present study omits multidimensional turbulence and instability, and does not include the proton-catalyzed $\alpha$ capture reactions $^{12}$C(p,$\gamma$)$^{13}$O($\alpha$,p)$^{16}$O. Multi-dimensional studies including a more detailed nuclear reaction network are needed to quantify more accurately the progenitor conditions which are feasible in the double-detonation mechanism (e.g., the thresholds to trigger different burning regimes).

One of our particular interests as implied by the present study is whether DDT could occur even in the double-detonation scenario. It may increase the chance for the lower mass envelope, and may change a view on our dynamical and nucleosynthesis behavior in the double-detonation scenario. Further tackling this issue will require high-resolution multi-dimensional simulations, which we plan to conduct in the future.

## 5. Acknowledgments

The numerical calculations were carried out on Yukawa-21 at YITP in Kyoto University. K.M. acknowledges support from the Japan Society for the Promotion of Science (JSPS) KAKENHI grant JP18H05223, JP20H00174, and JP20H04737.